\documentclass[]{aastex631}

\usepackage{rotating}
\usepackage{graphicx}
\usepackage{amssymb}
\usepackage{newtxtext,newtxmath}
\usepackage{color}
\usepackage[T1]{fontenc}
\usepackage[utf8]{inputenc}
\usepackage[T1]{fontenc}
\DeclareUnicodeCharacter{03C3}{}
\begin{document}

\title{Systematic Search and Study of Short-Timescale Flare Structures in BL Lac object Gamma-ray Emission}

\author{Jinjie Yu}
\affiliation{School of Physical Science and Technology, Kunming University, Kunming 650214, People’s Republic of China}

\author{Nan Ding}
\altaffiliation{Corresponding Author:\href{mailto:orient.dn@foxmail.com}{orient.dn@foxmail.com}}
\affiliation{School of Physical Science and Technology, Kunming University, Kunming 650214, People’s Republic of China}
\author{Junhui Fan}
\affiliation{Center for Astrophysics, Guangzhou University, Guangzhou 510006, China}
\affiliation{Astronomy Science and Technology Research Laboratory of Department of Education of Guangdong Province, Guangzhou 510006, China}
\author{Yunyong Tang}
\affiliation{School of Physical Science and Technology, Kunming University, Kunming 650214, People’s Republic of China}
\author{Jin Cao}
\affiliation{School of Physical Science and Technology, Kunming University, Kunming 650214, People’s Republic of China}



\begin{abstract}

We present here the first systematic search of short timescale $\gamma$-ray flares from 29 high Galactic latitude BL Lac objects over 14 years of Fermi Large Area Telescope data. Using a combined Bayesian Blocks and HOP algorithm, we identified seven high-quality orbital timescale flare segments from three sources and quantified 24 short-timescale flare structures. We then performed a comprehensive analysis of flare symmetry, power spectral density (PSD) of variability, and flux-photon index relation. The main results are as follows: (1) The flare symmetry parameter $A$ shows a "U-shaped" distribution. Short timescale flares are symmetric while long timescale flares are asymmetric. The number of fast-rise slow-decay and slow-rise fast-decay type flares are equal. No correlation is found between $A$ and peak/integral flux. No parameter evolution is seen between consecutive flares either. The observations support a scenario where longer timescale flares originate from superposition of short, symmetric sub-hour flares. (2) PSD from yearly to hourly timescales is modeled using the CARMA process. At lower frequencies, the PSD follows the typical broken power-law form. The high-frequency region of the PSD exhibits a continuous power-law shape, indicating that $\gamma$-ray variability originates from a single physical process across all probed timescales. (3) The flux-photon index distribution shows a pattern of "harder-when-brighter" or "softer-when-brighter," but becomes flat above a certain critical flux, with $\Gamma$ $\approx$ 2. This behavior cannot be simply explained by a two-component or blazar sequence model, and we speculate it may be related to complex interplay between electron acceleration and cooling. 

\end{abstract}

\keywords{Blazars (164) --- BL Lacertae objects (158)}


\section{Introduction} \label{sec:intro}

Blazars are one of the most peculiar subclasses of active galactic nuclei (AGN), exhibiting significant variability/flare phenomena and radiation with high polarization (\citealt{blandford2019}; \citealt{hovatta2019}). Based on the line width of optical emission, blazars are divided into two subclasses: BL Lacertae objects (BL Lacs) and flat-spectrum radio quasars (FSRQs). BL Lacs typically exhibit no or weak emission features (rest-frame equivalent width, EW \textless 5 Å), while FSRQs are characterized by strong emission lines (EW \textgreater 5 Å, e.g., \citealt{scarpa1997}; \citealt{ghisellini2011}). Based on the peak frequency of their synchrotron emission, blazars can also be categorized as high-synchrotron-peaked blazars (HSP, $\nu_{\mathrm{s}} \textgreater 10^{15}$ Hz), intermediate-synchrotron-peaked blazars (ISP, $10^{14}$ Hz $\textless \nu_{\mathrm{s}} \textless 10^{15}$ Hz), and low-synchrotron-peaked blazars (LSP, $\nu_{\mathrm{s}} \textless 10^{14}$ Hz, see e.g., \citealt{abdo2010}; \citealt{fan2016}; \citealt{yang2022}).

The spectral energy distributions (SEDs) of blazars typically exhibit a dual-peak structure, with a low-energy peak (in the radio to soft X-ray bands) and a high-energy peak (in the X-ray to TeV $\gamma$-ray bands). The radiation of the low-energy peak is commonly attributed to synchrotron emission produced by relativistic electrons. However, the origin of the high-energy peak is still a matter of debate. In the lepton model, the high-energy radiation is generated through the inverse Compton (IC) scattering of low-energy photons by relativistic electrons. There are several possible sources of those low-energy seed photons, including produced by synchrotron radiation from relativistic electrons (i.e., Synchrotron Self-Compton scenario, see e.g., \citealt{schlickeiser2009}; \citealt{niedzwiecki2012}) or from external photon fields, such as accretion disks, broad-line regions, and dusty torus (External Compton scenario, see e.g., \citealt{dermer1992}; \citealt{sikora1994}; \citealt{ghisellini2009can}). In the hadronic model, the high-energy radiation is attributed to either proton synchrotron emission or synchrotron radiation from secondary charged particles produced through strong interactions of protons (\citealt{mannheim1992}; \citealt{aharonian2000}; \citealt{bottcher2007}). The distinctive persistent non-thermal radiation of blazars makes them ideal objects for studying the acceleration and radiation of high-energy particles in jets. 

Another characteristic of blazars is the presence of significant variability across the entire electromagnetic spectrum. Most surprisingly, high-energy flares with short timescales have been detected in multiple sources, with flares occurring on timescales of hours or even minutes (e.g., \citealt{aharonian2007}; \citealt{ackermann2016}; \citealt{raiteri2017}; \citealt{wang2022}). The extremely short-term variability suggests a highly compact emission region and poses a serious challenge to traditional radiation models (see \citealt{bottcher2007}; \citealt{ghisellini2008}). Furthermore, compared to long-term variability, the physical mechanisms triggering these rapid variability/flare are expected to be relatively "simple", which provides a relatively "clean" event for studying particle acceleration processes within the jet. We note that systematic searches for the rapid variability/flare of blazars using an objective method and sample statistical studies on the observational characteristics of the rapid variability/flare are still lacking. Most of the current research is focused on individual sources (see e.g., \citealt{gaidos1996}; \citealt{catanese2000}; \citealt{aharonian2002}; \citealt{aharonian2007}; \citealt{albert2007}; \citealt{ghisellini2009}; \citealt{ghisellini2009tev}; \citealt{biteau2011}; \citealt{blinov2011}; \citealt{arlen2012}; \citealt{donnarumma2019}; \citealt{rulten2022}), and the varying criteria for defining the rapid flare across different works hampers the possibility of conducting sample analyses. \cite{meyer2019} conducted a comprehensive search and analysis of short-timescale $\gamma$-ray flares in six FSRQs based on nearly 10 years of Fermi-LAT data. They employed an objective method combining Bayesian block and HOB algorithm to identify the short-timescale flare structures. They discovered sub-hour flares in two sources, 3C 279 and CTA 102, and systematically analyzed the asymmetry of flare profiles, potential absorption features of $\gamma$-ray spectra by the broad-line region (BLR), and the time-delay correlation between $\gamma$-ray and radio/millimeter light curves. This study provides the first systematic observational analysis of rapid flares in FSRQs. 

A common view holds that FSRQs and BL Lac objects have drastically different central engine environments. FSRQs tend to possess more powerful jet power and higher accretion rates, while BL Lac objects exhibit the opposite, resulting in significant distinctions in their spectra and radiation mechanisms (see e.g., \citealt{blandford2019}). However, there is a lack of relevant research on whether there are differences between FSRQs and BL Lacs in terms of short-term flare phenomena. Compared to FSRQs, BL Lac objects have a relatively simpler radiation mechanism dominated by Synchrotron Self-Compton. In this context, short-term flare events in BL Lacs can serve as ideal targets for studying particle acceleration mechanisms. This is because by inferring the evolution of the electron spectrum through SED, one can greatly mitigate the uncertainties introduced by the radiation model dependence. This provides a feasible avenue for deepening our understanding of the triggering mechanisms behind short-term variability/flaring events. 

Motivated by those, in this paper, we systematically search for short-timescale flare structures in 29 BL Lacs located at high galactic latitudes ($ \lvert b \rvert $ \textgreater 10 ) and with a flux significant parameter {Signif\_Avg} \textgreater 100 based on nearly 14 years of observation data from Fermi-LAT. We employed an objective approach to identify these structures and analyzed their symmetry, power spectral density functions (PSD), as well as the relation between flux and photon index. The main structure is as follows: Section ~\ref{sec:searc} introduces the search sample, search strategy, and results. Section ~\ref{sec:symme} presents the symmetry analyses. Section ~\ref{sec:power} presents the PSD analyses. The relation analyses between flux and photon index are presented in Section ~\ref{sec:flux}. A summary is given in Section ~\ref{sec:Summary}.

\section{The search for short-timescale flare structures} \label{sec:searc}

To systematically search for hour/sub-hour rapid flare structures in BL Lacs, we first constructed a search sample. The sample was constructed based on the selection of sources from the high Galactic latitude ($ \lvert b \rvert $ \textgreater 10 ) sources of the Fermi LAT Fourth AGN Catalog. The reason for selecting the high Galactic latitude source is to minimize the impact of $\gamma$-ray radiation from the Galactic disk on the light curve of the sources, especially at short timescales. The criteria for selecting the sample source are as follows. 

(1) the flux significance parameter, {Signif\_Avg}, of the source is greater than 100. 

(2) the analysis flags parameter, "Flags", of the source is equal to 0.

\begin{table}[]
\caption{The information of the search sample}
\begin{center}
\begin{tabular}{lcccccccc}
\hline
\multicolumn{1}{c}{Source Name} & RA    & DEC   & Redshift & {Signif\_Avg} & Flux1000 & Unc\_{Flux1000} & Frac\_{Variability} & SED Classification \\
\multicolumn{1}{c}{}            &         &         &            &              & (10$^{-8}$ph cm$^{-2}$s$^{-1}$)        & (10$^{-10}$ph cm$^{-2}$s$^{-1}$)           &                             &         \\
\multicolumn{1}{c}{(1)}         & (2)     & (3)     & (4)        & (5)          & (6)              & (7)                  & (8)                & (9)                  \\ \hline
4FGL J0112.1+2245               & 18.029  & 22.752  & 0.265      & 140.416      & 0.84             & 1.37                 & 0.276              & ISP                  \\
4FGL J0144.6+2705               & 26.15   & 27.09   & -          & 110.253      & 0.62             & 1.2                  & 0.269              & ISP                  \\
4FGL J0211.2+1051               & 32.809  & 10.857  & 0.2        & 103.421      & 0.62             & 1.26                 & 0.407              & ISP                  \\
4FGL J0222.6+4302               & 35.67   & 43.036  & 0.444      & 182.312      & 1.39             & 1.78                 & 0.434              & ISP                  \\
4FGL J0238.6+1637               & 39.668  & 16.618  & 0.94       & 135.105      & 0.979            & 1.55                 & 1.083              & LSP                  \\
4FGL J0303.4-2407               & 45.863  & -24.123 & 0.266      & 107.544      & 0.479            & 1.05                 & 0.432              & HSP                  \\
4FGL J0428.6-3756               & 67.173  & -37.94  & 1.11       & 256.684      & 2.14             & 2.13                 & 0.5                & LSP                  \\
4FGL J0449.4-4350               & 72.358  & -43.835 & 0.205      & 166.501      & 1.25             & 1.66                 & 0.295              & HSP                  \\
4FGL J0509.4+0542               & 77.359  & 5.701   & 0.3365     & 113.378      & 0.816            & 1.53                 & 0.602              & ISP                  \\
4FGL J0538.8-4405               & 84.709  & -44.086 & 0.892      & 234.751      & 1.66             & 1.81                 & 0.802              & LSP                  \\
4FGL J0721.9+7120               & 110.488 & 71.341  & 0.127      & 321.002      & 2.37             & 1.93                 & 0.312              & ISP                  \\
4FGL J0818.2+4222               & 124.557 & 42.382  & 0.53       & 116.4        & 0.582            & 1.11                 & 0.256              & LSP                  \\
4FGL J0854.8+2006               & 133.707 & 20.116  & 0.306      & 103.384      & 0.533            & 1.1                  & 0.515              & LSP                  \\
4FGL J0958.7+6534               & 149.69  & 65.568  & 0.367      & 105.123      & 0.414            & 0.825                & 0.798              & ISP                  \\
4FGL J1015.0+4926               & 153.768 & 49.434  & 0.212      & 169.367      & 0.817            & 1.23                 & 0.236              & HSP                  \\
4FGL J1104.4+3812               & 166.119 & 38.207  & 0.03       & 343.769      & 3.42             & 2.74                 & 0.281              & HSP                  \\
4FGL J1217.9+3007               & 184.476 & 30.118  & 0.13       & 145.88       & 0.981            & 1.53                 & 0.331              & HSP                  \\
4FGL J1427.0+2348               & 216.756 & 23.801  & 0.6035     & 163.711      & 1.11             & 1.59                 & 0.251              & HSP                  \\
4FGL J1517.7-2422               & 229.425 & -24.373 & 0.048      & 101.836      & 0.667            & 1.33                 & 0.181              & ISP                  \\
4FGL J1555.7+1111               & 238.931 & 11.188  & 0.36       & 120.44       & 1.48             & 2.39                 & 0.177              & HSP                  \\
4FGL J1653.8+3945               & 253.474 & 39.76   & 0.033      & 173.434      & 1.01             & 1.42                 & 0.332              & HSP                  \\
4FGL J1748.6+7005               & 267.158 & 70.097  & 0.77       & 113.422      & 0.464            & 0.881                & 0.392              & ISP                  \\
4FGL J1800.6+7828               & 270.173 & 78.467  & 0.68       & 144          & 0.579            & 0.914                & 0.408              & ISP                  \\
4FGL J1903.2+5540               & 285.808 & 55.677  & -          & 100.962      & 0.432            & 0.903                & 0.177              & ISP                  \\
4FGL J2000.0+6508               & 300.011 & 65.148  & 0.047      & 168.944      & 0.958            & 1.31                 & 0.525              & HSP                  \\
4FGL J2139.4-4235               & 324.855 & -42.59  & -          & 107.75       & 0.594            & 1.19                 & 0.497              & ISP                  \\
4FGL J2158.8-3013               & 329.714 & -30.225 & 0.116      & 238.6        & 2                & 2.19                 & 0.252              & HSP                  \\
4FGL J2202.7+4216               & 330.695 & 42.282  & 0.069      & 239.123      & 2.65             & 2.45                 & 0.434              & ISP                  \\
4FGL J2236.5-1433               & 339.144 & -14.556 & 0.325      & 106.063      & 0.574            & 1.2                  & 0.777              & LSP                  \\ \hline             
\end{tabular}
\end{center}
\tablecomments{Columns 5 to 8 are extracted from Table 12 of \cite{abdollahi2020}. {Signif\_Avg} represents the significance of the source in $\sigma$ units over the 100 MeV to 1 TeV energy range. Flux1000 indicates the integrated photon flux from 1 to 100 GeV. Unc\_{Flux1000} represents the 1 $\sigma$ error on the integrated photon flux. Frac\_{Variability} is the fractional variability calculated based on the fluxes observed in each year. Column 9 is a categorization of source SEDs, taken from \cite{yang2022}.}
\label{table1}
\end{table}

\begin{figure}[h]
\centering
\includegraphics[width=0.45\textwidth]{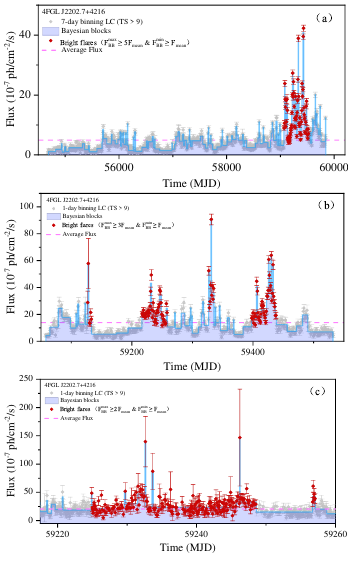} 
\caption{An example illustrating the iterative process of searching for bright flare regions with shorter timescales. The gray dots represent the data points of the light curve, the red dots indicate the identified bright flare region data points, the pink dashed line represents the average flux, and the blue solid line represents the Bayesian blocks. Please refer to Section ~\ref{sec:searc} for a detailed description of the iterative process. Figures (a), (b), and (c) respectively show the 7-day binning light curve of the source 4FGLJ2202.7+4216, the 1-day binning light curve obtained by zooming in on the bright flare region in Figure (a), and the orbit timescale binning light curve obtained by further zooming in on the second bright flare region in Figure (b). }
\label{fig1}
\end{figure}

\begin{figure}[h] 
\centering
\includegraphics[width=1\textwidth]{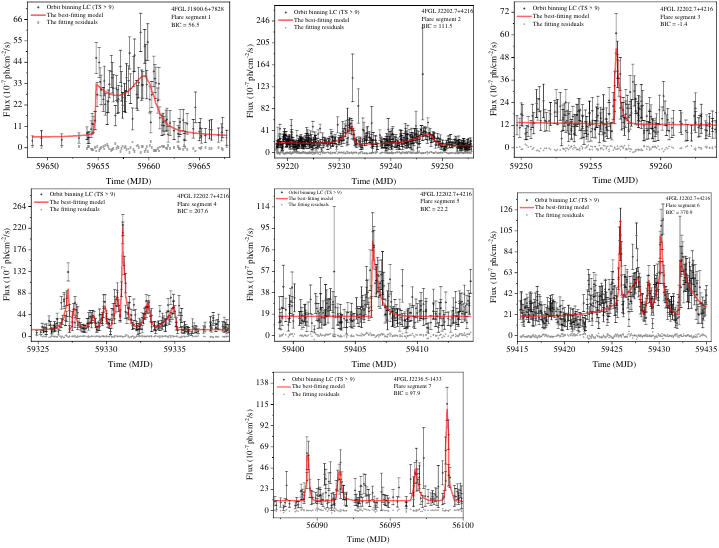}
\caption{Seven high-quality orbital timescale flare segments. The black dots represent the photometric data, while the red lines represent the best-fit quantitative results of short-time flare structures (see Section ~\ref{sec:symme}). The gray dots are the fitting residuals. The source name, flare segment ID, and the fitted BIC value are labeled in the top right corner. }
\label{fig2}
\end{figure}

Here the first criterion is to select the bright source, ensuring that they are sufficiently bright to have enough statistical significance on short timescales, and the second criterion is to ensure that the source is of high quality ("clean"). Finally, we obtain 29 sources as the final search sample from 1027 "clean" BL Lacs in the High Galactic Latitude Fermi LAT Fourth AGN Catalog (\citealt{abdollahi2020}). The details of the 29 sources are listed in Table ~\ref{table1}. It is worth noting that some evidence/suggestions indicate that BL Lac classification may be contaminated by FSRQ due to the jet continuum swamping the emission lines (\citealt{padovani2019}). In light of this, a more physically meaningful SED classification of those sources is also given in Table ~\ref{table1}. Out of the 29 sources, 6 (21\%) are LSP, 13 (45\%) are ISP, and 10 (34\%) are HSP. Some of these LSP and ISP sources might essentially be FSRQs, and their radiation mechanisms will deviate from the simple SSC scenario. We retrieved the latest P8R3 data \footnote{https://fermi.gsfc.nasa.gov/ssc/data/} of the 29 sources from the Fermi data server, spanning from August 4, 2008, to October 1, 2022.
In order to search for short-timescale bright flare structures, we employed an objective approach similar to the joint Bayesian block (BB) and HOP algorithm proposed by \cite{meyer2019} to perform iterative search. Specifically, we analyzed the data following the Fermi data processing standard procedures using Fermi Science Tools (v11r5p3) \footnote{https://fermi.gsfc.nasa.gov/ssc/data/analysis/software/v11r5p3.html} and the latest \texttt{P8R3\_SOURCE\_V3} \footnote{https://fermi.gsfc.nasa.gov/ssc/data/access/lat/BackgroundModels.html} instrument response functions. Detailed data selection criteria and processing steps can be found in Section ~\ref{sec:searc} of \cite{ding2019}. To prepare for handling variability at different time scales, a binned likelihood analysis to the total acquired data is first performed with gtlike to obtain an initial simple power-law (PL) form spectral model. Based on the initial spectral model, a 7-day binned light curve is generated (see Figure ~\ref{fig1}a for an example). We adopted a joint BB and HOP algorithm proposed by \cite{meyer2019} to iteratively search for short-time-scale bright flare regions. Specifically, we first utilized the Bayesian block (BB) algorithm (\citealt{scargle2013}) to obtain the optimal step function representation of the 7-day binning light curve. The BB algorithm is an adaptive data segmentation algorithm that automatically determines the optimal number of blocks and block boundaries by maximizing the goodness of fit, enabling quantitative characterization of time series with discontinuous changes such as mutations and outbursts. Then, the HOP algorithm was used to identify the flare regions. The specific operation is to first identify the block whose flux is higher than both the previous and the next block and satisfies the condition $F_{\mathrm{BB}} \geq F_{\mathrm{max}}$ as the peak block. Traveling downward from this peak block in both left and right directions, the movement continues as long as the adjacent blocks are consecutively lower and meet the condition $F_{\mathrm{BB}} \geq F_{\mathrm{mean}}$. Ultimately, the range that fulfill these criteria are identified as a HOP group (i.e., bright flare regions). If there are overlapping HOP groups, they are merged into a single region. Using $F_{\mathrm{max}}=5 \times F_{\mathrm{mean}}$ as the peak condition for HOP (consistent with the criteria used by \citealt{meyer2019}), bright flare regions in the 7-day binning light curve are identified (red data points in Figure ~\ref{fig1}a). These flare regions are further used to generate a 1-day binning light curve (as shown in Figure ~\ref{fig1}b). Similarly, applying the BB and HOP algorithms to the 1-day binning light curve, using $F_{\mathrm{max}}=3 \times F_{\mathrm{mean}}$ as the peak condition for HOP, bright flare regions are re-identified and further zoomed in to generate an orbital-timescale (~95 min) binning light curve (as shown in Figure ~\ref{fig1}c). After experimentation, we find that only the bright flare regions with an average flux greater than 2 $\times$ 10$^{-6}$ ph cm$^{-2}$ s$^{-1}$ could obtain high-quality orbital-timescale light curves. Therefore, we ultimately retained only the high-quality orbital timescale light curves generated from regions with $F_{\mathrm{HOP}} \textgreater$ 2 $\times$ 10$^{-6}$ ph cm$^{-2}$ s$^{-1}$ for further analysis. Through the aforementioned search process, we ultimately identified seven high-quality orbital timescale flare segments (using $F_{\mathrm{max}}=2 \times F_{\mathrm{mean}}$ as the HOP peak condition for the orbital-timescale binning light curve) in 3 sources (4FGLJ1800.6+7828, 4FGLJ2202.7+4216, and 4FGLJ2236.5-1433) out of the 29 sources, as shown in Figure ~\ref{fig2}.

\section{Symmetry Analysis} \label{sec:symme}

There is currently no consensus on the symmetry properties of the time profiles of flares in blazars, as well as the physical mechanisms driving these flares. In this section, we performed quantitative fitting for the short-timescale flare structures in the orbital timescale flare segments, and investigate the symmetry features of them. We employed the following commonly used mathematical form to quantify the flare structures in the seven orbital timescale light curves:\begin{equation}F_{\mathrm{flare}}(t)=\sum_{i=1}^{N_{\mathrm{flare}}}F_{\mathrm{flare},i}(t)+F_{\mathrm{bkg}}(t)\end{equation}where $F_{\mathrm{flare}, i}$ represents the i-th flare, with the specific form as follows:
\begin{equation}F_{\text {flare }, i}(t)= F_{0} \times\left[\exp \left(\frac{t-t_{0}}{\tau_{\text {rise }}}\right)+\exp \left(\frac{t_{0}-t}{\tau_{\text {decay }}}\right)\right]^{-1}\end{equation}$\tau_{\text{decay}}$ represents the decay timescale; $\tau_{\text{rise}}$ denotes the rise time scale; $t_{0}$ is the peak time of the flare, and $F_{0}$ represents half of the peak flux of the flare. $F_{\mathrm{bkg}}(t)$ represents the background flux, which is characterized by a simple linear function with a slope of $\beta$ and an intercept of $b$. We adopt an algorithm based on iterative searching for the minimum Bayesian information criterion (BIC) to determine the optimal number of flares and initial fitting parameters for each light curve segment. The BIC is defined as $\mathrm{BIC}=n_{\mathrm{par}} \ln (n)+\chi^2$. Here $n_\mathrm{par}$ represents the number of fitting parameters, $n$ is the number of data points, and $\chi^2$ is the fitting chi-square. The advantage of BIC lies in its ability to penalize the number of model parameters when evaluating model fit, thereby avoiding overfitting issues. After determining the optimal number of flares and initial model parameters, we finalized the best-fit parameters and their 1 $\sigma$ uncertainties of each flare using the Markov Chain Monte Carlo (MCMC) method. Among the seven orbital timescale light curves, we finally identified 24 flare structures (see Figure ~\ref{fig2}), and the best-fit parameters of each flare are listed in Table ~\ref{table2}. Table ~\ref{table2} also includes the calculated results of the symmetry parameter $A$ ($A= \left(\frac{\tau_{\text {rise }}-\tau_{\text {decay }}}{\tau_{\text {rise }}+\tau_{\text {decay }}}\right)$), the duration time $T_{\mathrm{90}}$ (the duration corresponding to 90\% of the integrated flux), and the integrated flux parameter $F_{\mathrm{int}}$ for each flare. It should be noted that, during the fitting process, we only considered data points with TS $\geq 9$. The ignored data points with TS $\textless 9$ account for an average of $\sim25\%$ of the entire light curve. These ignored low-luminosity data points would significantly affect the estimation of the background flux parameters $\alpha$ and $\beta$, but have minimal impact on the timescale parameters of flare structures. Subsequent work will focus on the timescale of flare structures. Therefore, the omission of these data points will not have a substantial impact on the main results of this paper.

\begin{sidewaystable}
\caption{Flare structure parameters }
\begin{center}
\scalebox{0.75}{
\begin{tabular}{lccccccccccc}
\hline
Source Name          & Flare segment ID     & Flare ID             & $\beta$              & $b$                  & $F_{0}$                                               & $t_{0}$                & $\tau_{\text{rise}}$ & $\tau_{\text{decay}}$ & $A$                             & $T_{90}$ & $F_{int}$                                             \\
\multicolumn{1}{l}{} & \multicolumn{1}{l}{} & \multicolumn{1}{l}{} & \multicolumn{1}{l}{} & \multicolumn{1}{l}{} & (10$^{-7}$ph cm$^{-2}$s$^{-1}$) & (day)                  & (day)                & (day)                 & \multicolumn{1}{l}{}            & (day)    & (10$^{-7}$ph cm$^{-2}$s$^{-1}$) \\
(1)                  & (2)                  & (3)                  & (4)                  & (5)                  & (6)                                                   & (7)                    & (8)                  & (9)                   & (10)                            & (11)     & (12)                                                  \\ \hline
4FGLJ1800.6+7828 & 1                & 1        & $0.001\pm0.001$   & $0.625\pm0.268$     & $41.932\pm12.723$ & $59659.905\pm15245.749$ & $1.617\pm0.505$      & $0.593\pm0.239$        & $0.464\pm0.201$                 & 7.079    & 76.511    \\
                 &                  & 2        & $-$               & $-$                 & $27.344\pm7.213$  & $59654.720\pm1.569$     & $0.008\pm0.002$      & $4.121\pm1.955$       & $-0.997\pm0.003$                & 146.819   & 112.531   \\
4FGLJ2202.7+4216 & 2                & 1        & $-0.155\pm0.012$ & $9173.801\pm705.508$ & $45.912\pm2.781$  & $59232.722\pm3.434$    & $1.502\pm0.114$      & $0.177\pm0.039$       & $0.789\pm0.044$                  & 6.862    & 70.225    \\
                 &                  & 2        & $-$               & $-$                 & $32.369\pm1.759$   & $59247.964\pm4.279$    & $3.182\pm0.257$      & $0.615\pm0.077$       & $0.677\pm0.041$                 & 14.375   & 107.565   \\
4FGLJ2202.7+4216 & 3                & 1        & $-0.091\pm0.065$ & $5350.656\pm3856.715$ & $58.589\pm6.437$  & $59256.743\pm0.017$    & $0.036\pm0.013$      & $0.248\pm0.034$       & $-0.748\pm0.081$                & 1.125    & 14.862    \\
4FGLJ2202.7+4216 & 4                & 1        & $-0.001\pm0.001$  & $18.639\pm0.002$     & $396.022\pm0.041$ & $59331.142\pm6.362$    & $0.039\pm0.001$       & $0.123\pm0.001$       & $-0.517\pm0.001$                & 0.543    & 53.694    \\
                 &                  & 2        & $-$               & $-$                 & $108.308\pm0.011$ & $59332.949\pm6.148$     & $0.185\pm0.001$      & $0.165\pm0.001$       & $0.059\pm0.001$                 & 0.889    & 29.647    \\
                 &                  & 3        & $-$               & $-$                 & $139.305\pm0.015$ & $59330.741\pm5.815$    & $0.175\pm0.001$      & $0.096\pm0.001$       & $0.295\pm0.001$                 & 0.759    & 30.138    \\
                 &                  & 4        & $-$               & $-$                 & $88.378\pm0.009$  & $59329.856\pm5.509$    & $0.208\pm0.001$      & $0.087\pm0.001$       & $0.414\pm0.001$                 & 0.903    & 21.222    \\
                 &                  & 5        & $-$               & $-$                 & $103.909\pm0.009$  & $59327.236\pm6.899$    & $0.278\pm0.001$      & $0.004\pm0.001$       & $0.979\pm0.001$                 & 1.276    & 28.792    \\
                 &                  & 6        & $-$               & $-$                 & $68.389\pm0.009$   & $59327.614\pm5.478$    & $0.041\pm0.001$      & $0.221\pm0.001$       & $-0.688\pm0.001$                & 0.997    & 15.688    \\
                 &                  & 7        & $-$               & $-$                 & $43.995\pm0.004$  & $59329.191\pm6.681$    & $0.263\pm0.001$      & $0.054\pm0.001$       & $0.663\pm0.001$                 & 1.186    & 12.122    \\
                 &                  & 8        & $-$               & $-$                 & $53.849\pm0.006$   & $59331.559\pm5.603$    & $0.029\pm0.001$      & $0.296\pm0.001$       & $-0.826\pm0.001$                & 1.356    & 16.137    \\
4FGLJ2202.7+4216 & 5                & 1        & $-1.266\pm0.311$  & $75209.915\pm18488.589$ & $73.701\pm3.638$  & $59406.349\pm3.697$     & $0.001\pm0.012$      & $0.449\pm0.029$         & $-0.999\pm0.053$                    & 2.072    & 33.162    \\
4FGLJ2202.7+4216 & 6                & 1        & $-0.001\pm0.001$  & $6.079\pm0.001$      & $68.548\pm0.008$  & $59429.476\pm7.733$    & $0.069\pm0.001$      & $3.389\pm0.001$        & $-0.961\pm0.001$                & 15.606   & 232.509    \\
                 &                  & 2        & $-$               & $-$                 & $52.353\pm0.006$  & $59432.491\pm5.969$    & $0.003\pm0.001$      & $4.966\pm0.001$       & $-0.999\pm0.001$                    & 22.867   & 259.960    \\
                 &                  & 3        & $-$               & $-$                 & $159.271\pm0.018$ & $59425.904\pm5.829$    & $0.072\pm0.001$      & $0.072\pm0.001$       & $-0.004\pm0.001$                & 0.364    & 17.912    \\
                 &                  & 4        & $-$               & $-$                 & $63.007\pm0.008$  & $59428.028\pm5.946$    & $9.081\pm0.001$      & $0.118\pm0.001$       & $0.975\pm0.001$                 & 41.811   & 572.266   \\
                 &                  & 5        & $-$               & $-$                 & $99.977\pm0.009$   & $59429.179\pm8.252$     & $0.589\pm0.001$      & $0.133\pm0.001$       & $0.634\pm0.001$                 & 2.645    & 62.232   
                  \\
                 &                  & 6        & $-$               & $-$                 & $88.874\pm0.009$   & $59429.659\pm5.542$    & $0.235\pm0.001$      & $0.004\pm0.001$       & $0.969\pm0.001$                  & 1.078    & 20.821    \\
4FGLJ2236.5-1433 & 7                & 1        & $0.002\pm0.001$  & $-86.967\pm0.009$      & $194.802\pm0.022$  & $56098.841\pm6.214$    & $0.055\pm0.001$      & $0.084\pm0.001$        & $-0.206\pm0.001$                & 0.370   & 21.402    \\
                 &                  & 2        & $-$               & $-$                 & $56.999\pm0.006$  & $56096.902\pm5.231$    & $0.178\pm0.001$      & $0.099\pm0.001$       & $0.286\pm0.001$                 & 0.770     & 12.562    \\
                 &                  & 3        & $-$               & $-$                 & $109.072\pm0.013$ & $56089.409\pm4.661$    & $0.061\pm0.001$      & $0.081\pm0.001$       & $-0.144\pm0.001$ & 0.366    & 12.092   
                 \\
                 &                  & 4        & $-$               & $-$                 & $61.504\pm0.007$ & $56091.569\pm6.023$    & $0.069\pm0.001$      & $0.143\pm0.001$       & $-0.343\pm0.001$ & 10.549    & 0.619    \\ \hline
\end{tabular}
}
\end{center}

\tablecomments{Column (1) is the source name; Column (2) is the number of the orbital timescale flare segment; Column (3) is the number of the flare structure identified in the flare segment; Columns (4)-(12) represent the parameters of the flare structure, as detailed in the text.}
\label{table2}
\end{sidewaystable}

\begin{figure}[h] 
\centering
\includegraphics[width=0.7\textwidth]{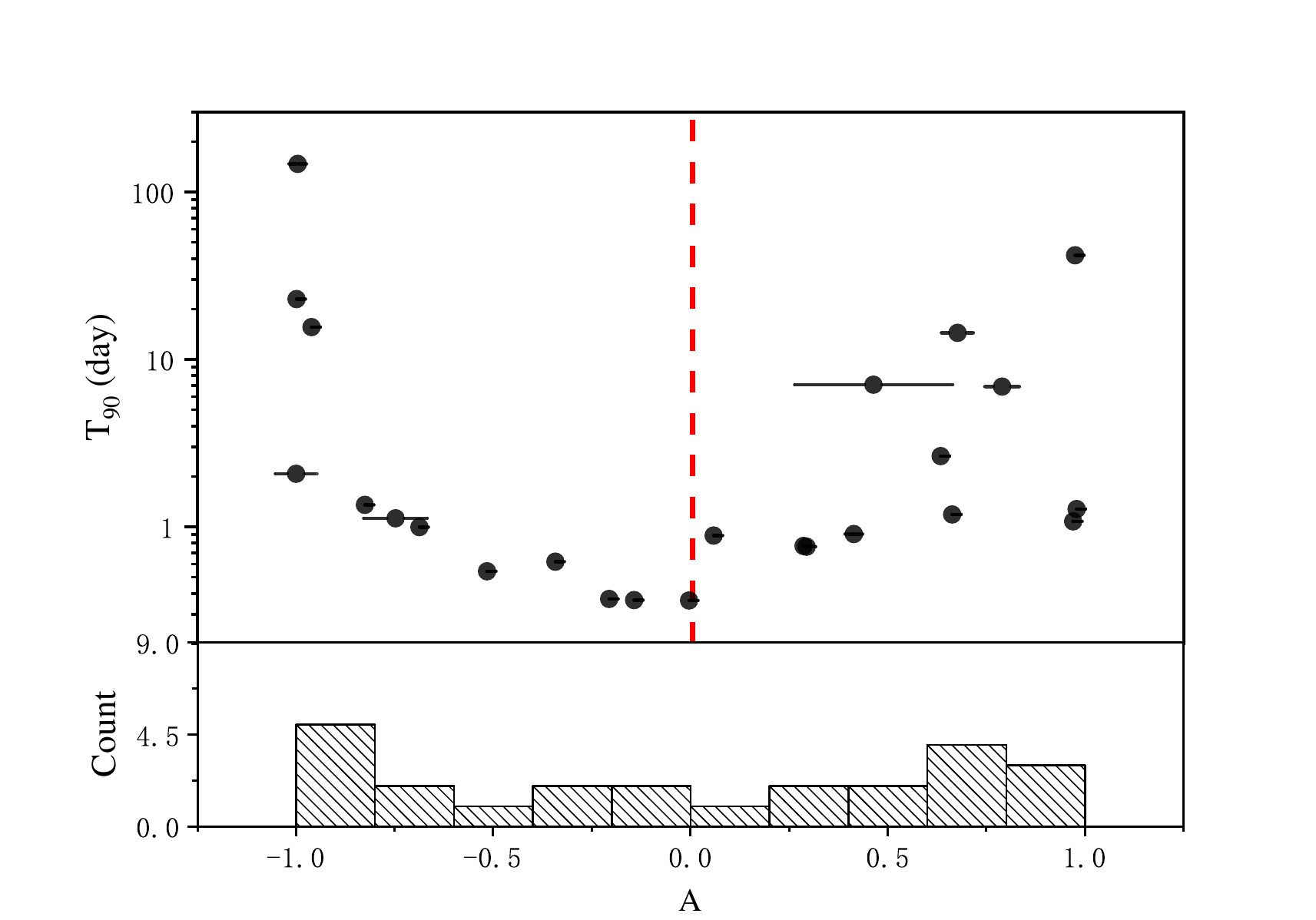}
\caption{The distribution of the symmetry parameter $A$ vs duration time $T_{\mathrm{90}}$.}
\label{fig3}
\end{figure}

The distribution of $A$ versus $T_{\mathrm{90}}$ is shown in Figure ~\ref{fig3}. Interestingly, at shorter timescales (on the order of a few hours), the flare profiles are symmetric. As the duration time increases (reaching the order of days to months), the flare profiles tend to exhibit pronounced asymmetry. Nevertheless, there is a similar proportion of fast rise slow decline (FRSD)-type flares and slow rise fast decline (SRFD)-type flares, with an average symmetry parameter of <A>=-0.01. Overall, the distribution of $A$ versus $T_{\mathrm{90}}$ exhibits a "U-shaped" trend (in fact, a similar result has emerged in Figure 7 of \citealt{meyer2019}). Moreover, similar to the results of \cite{meyer2019} for FSRQs, we find no correlation between the symmetry parameter $A$ and the flare half-peak flux parameter $F_{0}$, as well as the integral flux parameter $F_{\mathrm{int}}$ (Pearson correlation coefficients of -0.077 and 0.078, respectively). For the 4 th and 6 th orbital timescale light curves with multiple flare structures in Table ~\ref{table2}. Based on binomial distribution hypothesis testing, we analyzed whether there is an evolution in the peak flux, asymmetry, or duration of successive flares. Specifically, we calculated the p-value based on the assumption that the differences between the parameters of successive flares follow a binomial distribution where the probabilities of negative and positive differences are equal. The results, with p-values significantly greater than 0.05, indicate that there is no evidence of any evolution trends in the peak flux, asymmetry, or duration among successive flares. 

Some studies have shown that on long timescale, the temporal profiles of outbursts typically exhibit symmetric structures. \cite{chatterjee2012} conducted a two-year monitoring of six blazars and found that for the majority of long-term (timescale on several months) outburst profiles are symmetric in the GeV and optical bands. \cite{roy2019} further analyzed a larger sample of 10 blazars, comprising  $ \sim $200 long-term (timescale on several weeks to several months) outburst in the GeV and R-band, and confirmed this result. The symmetry of long-term outburst profiles is usually attributed to a simple explanation that the rise and decay timescales are predominantly governed by the crossing time of the shock front through the radiation region. Compared to long-term outburst, short-term flares exhibit complexity in their temporal profiles. \cite{roy2019} also analyzed the symmetry of  $ \sim $25 short-term (timescale on a few hours to several days) GeV flares in their sample and found that a significant proportion of short-term flares are asymmetric, but with similar FRSD-type flare and SRFD-type flare percentages. \cite{meyer2019} analyzed the short-term flares in six bright FSRQs and found that the majority of flares exhibited asymmetry, but the degree of asymmetry varies among each source and was not correlated with physical parameters such as luminosity. Moreover, they found that FRSD-type flares seem tend to be more common than SRFD-type flares (similar results have also been reported in some studies focusing on individual sources, e.g., \citealt{nalewajko2013}; \citealt{li2016}; \citealt{li2018}). The asymmetry in the temporal profiles of short-term flares is believed to be the result of a competition between electron acceleration and cooling timescales. However, \cite{nalewajko2013} proposed an alternative explanation, suggesting that the asymmetry could be attributed to the superposition of multiple individual flares with extremely short timescales from different emission regions (similar views are also discussed in \citealt{saito2013}). Our results confirm that on short timescales, flares exhibit diverse symmetric structures. Flares with durations on the scale of days to months show pronounced asymmetry, without a preference for FRSD-type or SRFD-type flares. It is noteworthy to note that for extremely short timescale (near hourly scales) flares tend to have symmetric structures. Moreover, there is no discernible evolutionary trend for a series of successive flares, and the symmetry of flares is also not correlated with parameters such as luminosity. This strongly suggests that each flare event is independent and has no direct correlation with each other. Such observational results are more consistent with the scenario where short timescale flares on the order of days to months are a result of the superposition of extremely short timescale flares on the hour scale with symmetric structures. In particular, the randomness of the superposition process can naturally explain the lack of a significant preference for FRSD-type or SRFD-type asymmetry in the flares on the day-month scale, without the need to introduce complex mechanisms in the competition for acceleration and cooling of electrons.

\cite{meyer2019} systematically searched for short-term flares in six FSRQs and ultimately discovered short-timescale flare structures in four sources (3C 279, CTA 102, PKS 1510–089, and 3C 454.3). The SED types of these four sources are all classified as LSPs. In our work, we identified high-quality orbital flare segments in three out of 29 sources and confirmed the presence of short-timescale flare structures in all three sources through fitting of the flare structures. Among these three sources, two are classified as ISPs and one as an LSP. Considering the results from \cite{meyer2019}, it currently appears that short-timescale flare tends to occur in sources with lower synchrotron peak frequencies. There are two possible reasons for this phenomenon: 1) For LSP/ISP sources (especially FSRQs), external photon fields (such as those from the accretion disk and broad-line region) may have a significant impact on the cooling process of radiation. Changes in these external photon fields lead to rapid variations in jet radiation. HSP sources are mainly dominated by the SSC process and are less affected by changes in external photon fields. 2) The complex external environments of LSP/ISP sources are more likely to lead to processes such as magnetic reconnection and plasma instabilities. These instabilities can change the physical conditions of the jet on short timescales, triggering rapid variability. In contrast, the jets in HSP sources are relatively more stable.

\section{Power Spectral Density Function Analysis} \label{sec:power}
 
 In this section, based on the Continuous-time autoregressive moving-average (CARMA) model fitting technique proposed by \cite{kelly2014}, we model the long-term (weekly binning) and short-term (orbital binning) light curves of the three sources having orbital timescale light curves and obtained power spectral density (PSD) functions that span approximately five orders of magnitude, ranging from the annual to hourly timescales. The CARMA model uses the following continuous stochastic differential equation to describe the stationary time series: 
 \begin{equation}\frac{d^py(t)}{dt^p}+\alpha_{p-1}\frac{d^{p-1}y(t)}{dt^{p-1}}+...+\alpha_{0}y(t)=\beta_{q}\frac{d^q\epsilon(t)}{dt^q}+\beta_{q-1}\frac{d^{q-1}\epsilon(t)}{dt^{q-1}}+...+\epsilon(t)
 \end{equation},
 where $\epsilon(t)$ is a Gaussian noise process with variance $\sigma^2$ and zero mean. The autoregressive polynomial order is p, the moving average polynomial order is q, and the corresponding CARMA model is denoted as CARMA(p, q). $\alpha=(\alpha_{0},...,\alpha_{q})$ and $\beta=(\beta_{0},...,\beta_{q})$ are the autoregressive and moving average coefficients, respectively. The CARMA model fitting technique utilizes Bayesian inference to compute the model parameters for an actual light curves under a given (p, q) orders. These parameters are then used to directly calculate the underlying PSD (see \citealt{kelly2014}):
 \begin{equation}P(f)=\sigma^2\frac{|{\sum_{j=0}^{q}}\beta_{j}\left(2\pi if\right)^{j}|^2}{|{\sum_{k=0}^{p}}\alpha_{k}\left(2\pi if\right)^{k}|^2}\end{equation}This approach offers sufficient flexibility to capture high-order variability features and generate more accurate PSD estimates. In the specific modeling process, we employed a grid search method to determine the optimal (p, q) orders for fitting the observed data with the CARMA model. In the grid search process, the maximum order for p is set to 8, and the maximum order for q is set to (p-1). We used the Bayesian information criterion (BIC) to quantitatively assess the goodness of fit for different (p, q) orders. Here BIC is formally defined as 
\begin{equation}\mathrm{BIC}=k\ln (n)-2 \ln (\hat{L})\end{equation}.
Where, $k$ is the number of parameters of the model, $n$ is the sample size, $\hat{L}$ is is the maximum likelihood estimate of the model. The use of BIC is motivated by its stronger penalty on model complexity compared to the Akaike information criterion (AIC), which helps avoid high-order models that overfit and produce spurious structures in the PSD. Additionally, BIC incurs less computational cost compared to the Deviance Information Criterion (DIC). After determining the optimal (p, q) orders for the CARMA model, the corresponding PSD is computed through sampling. To facilitate the comparison between the PSD obtained from long-term light curves and orbital timescale light curves, the PSD is normalized based on the following formula (\citealt{vaughan2003}): 
\begin{equation}P_{\mathrm{Normal}}=\frac{2\Delta T}{N\overline{x}}P(f)\end{equation}where $\frac{2\Delta T}{\overline{x}N}$ is the normalization factor. $\Delta T$ is the average sampling time, $N$ is the number of data points in the light curve, and $\overline{x}$ is the average flux.

\begin{figure}[h]
\centering
\includegraphics[width=0.98\textwidth]{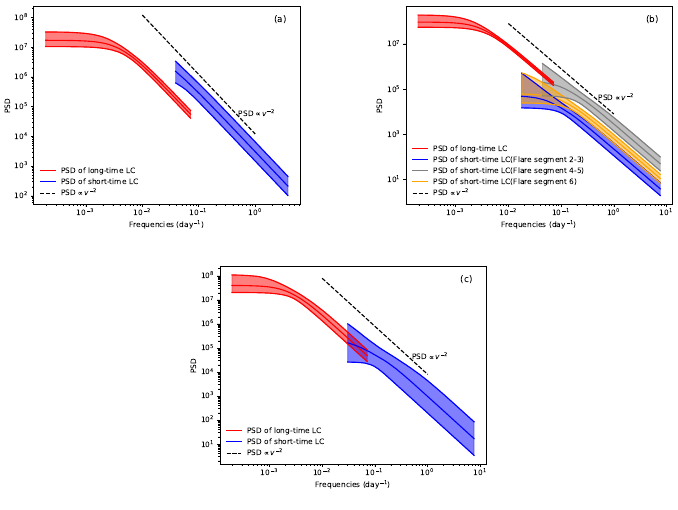} 
\caption{The PSDs of sources 4FGLJ1800.6+7828, 4FGLJ2202.7+4216, and 4FGLJ2236.5-1433. The red lines represent the PSDs obtained from the weekly binned light curve, while the blue, green, and yellow lines represent the PSDs obtained for the orbital timescale light curve. The corresponding color region represents the 1 $\sigma$ confidence interval. Additionally, the black dashed line represents a reference line indicating a power-law spectrum with a spectral index of -2. }
\label{fig4}
\end{figure}

Figure ~\ref{fig4} presents the PSDs calculated using the aforementioned method for the three BL Lacs. The red line is the PSD obtained from the weekly binned light curve, while the blue, green, and yellow lines represent the PSDs obtained for the orbital timescale light curve. The corresponding color region represents the 1$\sigma$ confidence interval. The black dashed line represents a reference line for a power-law spectrum with spectral index of -2. Thanks to the exceptional sky survey capabilities of Fermi-LAT, there have been several systematic analyses for the $\gamma$-ray variability of blazars based on Fermi-LAT data in recent years. These studies primarily focus on the power spectral density (PSD) in the frequency range of $10^{-4}$ to $10^{-1}$ days$^{-1}$, revealing that the PSD of blazars in the gamma-ray regime generally exhibits a break power law form with a break frequency between $10^{-3}$ to $10^{-2}$ days$^{-1}$. Above the break frequency, PSD follows a power law with a spectral index of $ \sim $-2. Below the break frequency, the PSD gradually transitions into a plateau. Here, the PSDs of the three sources show consistent results in the low-frequency regime ($10^{-4}-10^{-1}$ days$^{-1}$), confirming the previous findings. Moreover, it is worth noting that similar to previous results, the PSD break timescale is significantly longer than the timescales associated with particle radiation processes (e.g., electron cooling or acceleration time scales), but shorter than the typical thermal instability timescales associated with accretion processes. \cite{ruan2012} found that the break timescale of non-thermal radiation variability in BL Lacs at optical band is approximately four times smaller than that of normal quasars. They suggested that if thermal and non-thermal radiation variability are essentially of the same origin (i.e., thermal instability), the difference between the two timescales could be due to Doppler effect. \cite{zhang2022} further proposed that in addition to the Doppler effect, this discrepancy requires a difference in the relative position between the jet radiation region and the accretion disk radiation region to further alleviate it. We note, however, that the PSD simulation results of the variability driven by a series of continuous shocks with characteristic timescale of 100 or 150 day, which is reported by \cite{mukherjee2019}, show good agreement with the observed break timescale in here (see Figure 2 in \citealt{mukherjee2019}). Thus, this break may not be of origin associated with the accretion process, but rather with the dynamic processes of the continuous shock within the jet. 

Observational results of the PSD in the high-frequency regime are currently scarce. Only a few studies have reported interesting features in the PSD of individual sources in this regime. For instance, the blazar 3C 454.3 exhibits a prominent break in its normalized PSD at the frequency of $ \sim $1/7 day (\citealt{ackermann2010}; \citealt{nakagawa2013}; \citealt{ryan2019}). The PSDs of 3C 66A and PKS 2155-304 show break characteristics at the frequency of $ \sim $1/25 day and $ \sim $1/43 day, respectively (\citealt{sobolewska2014}). However, \cite{ryan2019} cautioned that the high-frequency break in the two sources may be false. These high-frequency breaks are believed to be associated with short-timescale internal shock processes (\citealt{nakagawa2013}) or radiation processes (e.g., electron escape, electron cooling; \citealt{finke2014}). They have distinct physical origins compared to the low-frequency breaks. Thanks to the availability of shorter timescale light curves, we extended the PSD to higher frequency ranges ($10^{-1}-10^{1}$ days$^{-1}$). Our results show that in the higher frequency regime, the normalized PSD follows the same power-law relation as in the low-frequency regime, without any apparent break. This result suggests that the variability of BL Lacs in the $\gamma$-ray band is most likely driven by a single (continuous) process.

\section{The relation of flux versus photon index} \label{sec:flux}

\begin{figure}[h] 
\centering
\includegraphics[width=0.9\textwidth]{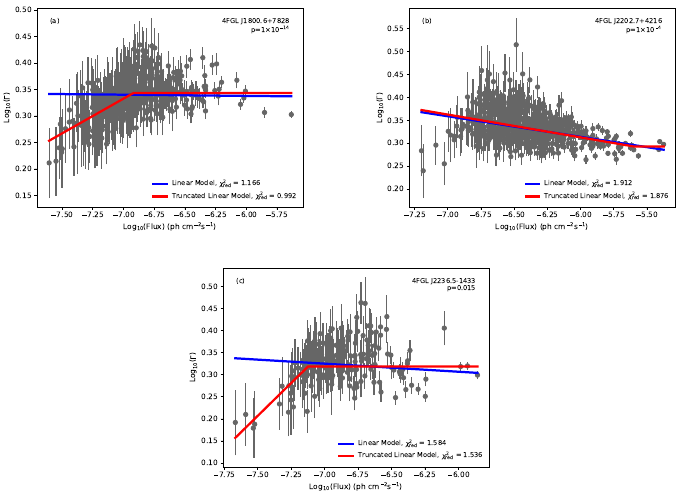} 
\caption{The flux versus photon index distribution in logarithmic coordinates. The green and red lines correspond to the fitted results of the linear and truncated linear models, respectively. The source names and the p-values of the T-test are labeled in the top right corner of each plane. }
\label{fig5}
\end{figure}

The relation of flux versus photon index can be used to trace the evolution of energy spectrum to reveal the physical mechanism of variability of blazars. Thus, Figure ~\ref{fig5} shows the flux versus photon index distribution in logarithmic coordinates for the three sources with orbital time-scale light curves (4FGLJ1800.6+7828, 4FGLJ2202.7+4216, and 4FGLJ2236.5-1433). From the figure, it can be found that: (1) in high-flux states, the dispersion of the photon index decreases, and the photon index converge to a constant of $\Gamma$ $ \sim $2; (2) in low-flux states, 4FGL J2236.5-1433 and 4FGL J1800.6+7828 exhibit a positive correlation between flux versus photon index (i.e., softening when brightening), while 4FGL J2202.7+4216 shows a weak inverse correlation (i.e., hardening when brightening). To quantitatively determine the confidence level of the truncation appearing in the flux-photon index plot from low to high flux states, we fitted the data using the following two different phenomenological models:

\begin{equation}\text{Model 1}:\Gamma=C \cdot f^{-a}\end{equation}
\begin{equation}\label{eq:mode2}
\begin{aligned}
\text{Model 2}: \Gamma = \begin{cases}
    C \cdot f^{-a} & \text{if } f < f_{b} \\
    C \cdot f^{-a}_{b} & \text{if } f \geq f_{b}
\end{cases}
\end{aligned}
\end{equation}

In logarithmic coordinates, Model 1 is a simple linear model, while Model 2 is a truncated linear model. In Model 2, above the truncation flux $f_{b}$, the photon index remains a constant of $C \cdot f^{-a}_{b}$, where $C$ is a proportional constant. We performed data fitting on the two models mentioned above using the Levenberg-Marquardt algorithm in Python lmfit module \footnote{\url{https://lmfit.github.io/lmfit-py/}}, taking into account the data errors. The best-fit results for each of the two models are shown in Figure ~\ref{fig5}, along with their corresponding reduced chi-square values. We conducted a T-test to assess the significance of the improvement in goodness-of-fit provided by the truncated linear model compared to the linear model. The p-value of the T-test is also displayed in Figure ~\ref{fig5}. From the statistical results ($p \ll 0.05$), the truncated linear model significantly outperforms the linear model in all three sources. This demonstrates that there are indeed truncated features in the flux versus photon index relations.

In the optical band, there is a strong correlation between the color (i.e., spectral index) and flux, and the correlation is associated with the type of blazars. Generally, BL Lacs exhibit a blue-when-brighter (BWB) behavior, while FSRQs show a red-when-brighter (RWB) behavior. However, in recent years, some studies noted that the BWB and RWB trends are not adequate to fully describe the flux-color relation in blazars (e.g., \citealt{yuan2017}; \citealt{fan2018}; \citealt{sarkar2019}; \citealt{safna2020}; \citealt{xiong2020}; \citealt{fang2022}), and they found that the color variations with flux follow more complex behaviors known as blue-stable-when-brighter (BSWB) or red-stable-when-brighter (RSWB), which is similar to the features appearing in Figure ~\ref{fig5} although here in the $\gamma$-ray band. The BWB trend is naturally expected from shock emission (\citealt{marscher1996}). However, the RWB behavior, or the presence of both BWB and RWB trends in a single source, requires additional explanation. One possible explanation is the interplay between accretion and jet radiation. \cite{zhangchu2022} proposed a hybrid model with double constant spectral index components to explain the behaviors of BSWB and RSWB observed in the optical band. In this model, the truncate is caused by the changing relative contributions of accretion disk thermal radiation and jet non-thermal radiation with luminosity (also see \citealt{zhang2023}). It is difficult, however, to straightforwardly generalize the model in the $\gamma$-ray band, where the radiation is completely dominated by non-thermal radiation from the jet. 
In the $\gamma$-ray band, the relation between photon index and flux has also been of wide interest. However, in the EGRET era, due to data quality limitations (spectral indices have large uncertainties), no clear results were established (\citealt{nandikotkur2007}). In the Fermi era, \cite{abdo2010a} noted that the photon index and gamma-ray flux generally show an inverse correlation in FSRQs and LSP BL Lacs, whereas this inverse correlation no longer exists in ISP and HSP BL Lacs. It is worth emphasizing that the phenomenon of the photon index evolving with increasing flux, showing an initial increase (or decrease) followed by a transition to a constant, is indeed widespread (see \citealt{bottcher2007the},\citealt{foschini2010},\citealt{acciari2023}), yet it has often been overlooked. In studies on samples such as Figure 5 in \cite{singal2012} it can also be found that for BL Lacs, the photon index softens with increasing flux and then converges to a constant of $\Gamma$ $ \sim $2 after a certain flux. The observed "BSWB" and "BSWB" behavior, similar to that found in optical band, is noteworthy, especially since the photon index stabilizes at a constant after a critical flux, and there must be an important physical mechanism behind this behavior. It is well known that the high-energy peak frequency of SED in blazars shifts to lower frequencies with increasing luminosity, which is called "blazar sequence". This can explain the observed trend of spectral softening with increasing flux in BL Lacs. However, it is incapable of explaining the observed anti-correlation between photon index and flux in FSRQs, as well as the behavior where the photon index converges to a constant value after reaching a certain critical luminosity/flux. We speculate that this behavior is more likely associated with the complex competition between electron acceleration and radiative cooling. It warrants a dedicated study in the future based on a large sample. 

\section{Summary} \label{sec:Summary}

In this work, based on nearly 14 years of Fermi-LAT observational data, we conducted a systematic search for short timescale flare segments for 29 BL Lacs located at high galactic latitudes with {Signif\_Avg} \textgreater 100 using an objective method of the combined BB and HOP algorithms. Among these objects, we successfully identified seven high-quality, orbital timescale flare segments in three sources (4FGLJ1800.6+7828, 4FGLJ2202.7+4216, and 4FGLJ2236.5-1433). Based on the long-term and short-term variability data, we conducted a comprehensive analysis of the symmetry of flares, the PSD of variability, and the relation between flux and photon index for the three sources. The main results are as follows:

1. Using a quantitative fitting technique, we identified 24 flare structures from seven orbital time-scale light curve segments. The distribution of the symmetry parameter $A$ versus the duration parameter $T_{\mathrm{90}}$ of these flares exhibits a pronounced "U-shaped" pattern. Specifically, flares characterized by shorter time scales (on the order of a few hours) display symmetrical profiles, while those with longer time scales (ranging from days to months) exhibit significantly asymmetrical profiles. Additionally, the proportion of the FRSD-type flare and the SRFD-type flare is equal, with no clear preference. There is no correlation between the symmetry parameter $A$ and the peak flux parameter $F_{0}$, as well as the integrated flux parameter $F_{\mathrm{int}}$. Moreover, there is no evidence for any evolution of parameters between continuous flare structures. These observational results are more consistent with a scenario where flare structures on timescales ranging from days to months are formed by the superposition of a series of symmetric, sub-hourly-scale short timescale flares. 

2. Using the CARMA model, we modeled the long-term and short-term light curves of the three sources that have orbital time-scale light curves, obtaining their PSD functions from annual to hourly scales ($ \sim $5 magnitudes). The results show that in the low-frequency regime, in agreement with previous results, the PSD exhibits a typical break power law form, and the characteristic timescale of the low-frequency break is longer than the timescale associated with particle radiation but shorter than the accretion-related thermal instability timescale. We note that this break timescale is perhaps not associated with the accretion process, but to the dynamic process of the continuous shock within the jet. Additionally, we extended the PSD to a higher frequency regime than previous studies. The results show that in the higher frequency regime, the PSD follows a power-law relation consistent with that of the lower frequency regime, with no obvious break features, which suggests that the $\gamma$-ray variability should be driven by a single (continuous) physical process. 

3. The flux and photon index distributions of the sources, 4FGLJ1800.6+7828, 4FGLJ2202.7+4216, and 4FGLJ2236.5-1433 exhibit behaviors similar to "BSWB" and "RSWB". Specifically, as the flux increases, the photon index initially shows a trend of becoming harder or softer, and then stabilizes at a constant value of $\Gamma$ $ \sim $2 after reaching a critical flux. This behavior cannot be explained simply by a two-component hybrid model of thermal and non-thermal radiation or by the "blazar sequence". We speculate that this is likely related to the complex competition between electron acceleration and radiative cooling processes. 

\section*{ACKNOWLEDGMENTS} \label{sec:acknow}
We sincerely thank the anonymous referee for constructive suggestions. We are grateful for the financial support from the National Natural Science Foundation of China (No. 12103022) and the Special Basic Cooperative Research Programs of Yunnan Provincial Under graduate Universities Association (No. 202101BA070001-043 and NO. 202301BA070001-104). N.D. is sincerely grateful for the financial support of the Xingdian Talents Support Program, Yunnan Province (NO. XDYC-QNRC-2022-0613). J.J.Y. sincerely appreciates the financial support from the Scientific Research Fund Program of Yunnan Provincial Department of Education (No. 2024Y751). J.H.Fan acknowledges the support from the NSFC U2031201. 

Facilities: Fermi (LAT).

Software: Fermi Science Tools (\citealt{2019ascl}), LMFIT (\citealt{newville2016}).

\bibliography{references}{}

\begin{thebibliography}{}
\expandafter\ifx\csname natexlab\endcsname\relax\def\natexlab#1{#1}\fi
\providecommand{\url}[1]{\href{#1}{#1}}
\providecommand{\dodoi}[1]{doi:~\href{http://doi.org/#1}{\nolinkurl{#1}}}
\providecommand{\doeprint}[1]{\href{http://ascl.net/#1}{\nolinkurl{http://ascl.net/#1}}}
\providecommand{\doarXiv}[1]{\href{https://arxiv.org/abs/#1}{\nolinkurl{https://arxiv.org/abs/#1}}}

\bibitem[{Abdo {et~al.}(2010{\natexlab{a}})Abdo, Ackermann, Agudo, Ajello, Aller, Aller, Angelakis, Arkharov, Axelsson, Bach, {et~al.}}]{abdo2010}
Abdo, A., Ackermann, M., Agudo, I., {et~al.} 2010{\natexlab{a}}, The Astrophysical Journal, 716, 30

\bibitem[{Abdo {et~al.}(2010{\natexlab{b}})Abdo, Ackermann, Ajello, Atwood, Axelsson, Baldini, Ballet, Barbiellini, Bastieri, Bechtol, {et~al.}}]{abdo2010a}
Abdo, A., Ackermann, M., Ajello, M., {et~al.} 2010{\natexlab{b}}, The Astrophysical Journal, 710, 1271

\bibitem[{Abdollahi {et~al.}(2020)Abdollahi, Acero, Ackermann, Ajello, Atwood, Axelsson, Baldini, Ballet, Barbiellini, Bastieri, {et~al.}}]{abdollahi2020}
Abdollahi, S., Acero, F., Ackermann, M., {et~al.} 2020, The Astrophysical Journal Supplement Series, 247, 33

\bibitem[{Acciari {et~al.}(2023)Acciari, Agudo, Aniello, Ansoldi, Antonelli, Engels, Artero, Asano, Baack, Babi{\'c}, {et~al.}}]{acciari2023}
Acciari, V., Agudo, I., Aniello, T., {et~al.} 2023, Astronomy \& Astrophysics, 670, A145

\bibitem[{Ackermann {et~al.}(2010)Ackermann, Ajello, Baldini, Ballet, Barbiellini, Bastieri, Bechtol, Bellazzini, Berenji, Blandford, {et~al.}}]{ackermann2010}
Ackermann, M., Ajello, M., Baldini, L., {et~al.} 2010, The Astrophysical Journal, 721, 1383

\bibitem[{Ackermann {et~al.}(2016)Ackermann, Anantua, Asano, Baldini, Barbiellini, Bastieri, Gonzalez, Bellazzini, Bissaldi, Blandford, {et~al.}}]{ackermann2016}
Ackermann, M., Anantua, R., Asano, K., {et~al.} 2016, The Astrophysical journal letters, 824, L20

\bibitem[{Aharonian(2000)}]{aharonian2000}
Aharonian, F. 2000, New Astronomy, 5, 377

\bibitem[{Aharonian {et~al.}(2002)Aharonian, Akhperjanian, Beilicke, Bernl{\"o}hr, B{\"o}rst, Bojahr, Bolz, Coarasa, Contreras, Cortina, {et~al.}}]{aharonian2002}
Aharonian, F., Akhperjanian, A., Beilicke, M., {et~al.} 2002, Astronomy \& Astrophysics, 393, L37

\bibitem[{Aharonian {et~al.}(2007)Aharonian, Akhperjanian, Bazer-Bachi, Behera, Beilicke, Benbow, Berge, Bernl{\"o}hr, Boisson, Bolz, {et~al.}}]{aharonian2007}
Aharonian, F., Akhperjanian, A., Bazer-Bachi, A., {et~al.} 2007, The Astrophysical Journal, 664, L71

\bibitem[{Albert {et~al.}(2007)Albert, Aliu, Anderhub, Antoranz, Armada, Baixeras, Barrio, Bartko, Bastieri, Becker, {et~al.}}]{albert2007}
Albert, J., Aliu, E., Anderhub, H., {et~al.} 2007, The Astrophysical Journal, 667, L21

\bibitem[{Arlen {et~al.}(2012)Arlen, Aune, Beilicke, Benbow, Bouvier, Buckley, Bugaev, Cesarini, Ciupik, Connolly, {et~al.}}]{arlen2012}
Arlen, T., Aune, T., Beilicke, M., {et~al.} 2012, The Astrophysical Journal, 762, 92

\bibitem[{Biteau \& Giebels(2011)}]{biteau2011}
Biteau, J., \& Giebels, B. 2011, in SF2A-2011: Proceedings of the Annual meeting of the French Society of Astronomy and Astrophysics, 525--528

\bibitem[{Blandford {et~al.}(2019)Blandford, Meier, \& Readhead}]{blandford2019}
Blandford, R., Meier, D., \& Readhead, A. 2019, Annual Review of Astronomy and Astrophysics, 57, 467

\bibitem[{Blinov {et~al.}(2011)Blinov, Hagen-Thorn, Hagen-Thorn, Takalo, \& Sillanp{\"a}{\"a}}]{blinov2011}
Blinov, D., Hagen-Thorn, V., Hagen-Thorn, E., Takalo, L., \& Sillanp{\"a}{\"a}, A. 2011, Astronomy reports, 55, 1078

\bibitem[{B{\"o}ttcher(2007)}]{bottcher2007}
B{\"o}ttcher, M. 2007, in The Multi-Messenger Approach to High-Energy Gamma-Ray Sources, Springer, 95--104

\bibitem[{B{\"o}ttcher {et~al.}(2007)B{\"o}ttcher, Basu, Joshi, Villata, Arai, Aryan, Asfandiyarov, Bach, Bachev, Berduygin, {et~al.}}]{bottcher2007the}
B{\"o}ttcher, M., Basu, S., Joshi, M., {et~al.} 2007, The Astrophysical Journal, 670, 968

\bibitem[{Catanese \& Sambruna(2000)}]{catanese2000}
Catanese, M., \& Sambruna, R.~M. 2000, The Astrophysical Journal, 534, L39

\bibitem[{Chatterjee {et~al.}(2012)Chatterjee, Bailyn, Bonning, Buxton, Coppi, Fossati, Isler, Maraschi, \& Urry}]{chatterjee2012}
Chatterjee, R., Bailyn, C., Bonning, E., {et~al.} 2012, The Astrophysical Journal, 749, 191

\bibitem[{Dermer {et~al.}(1992)Dermer, Schlickeiser, \& Mastichiadis}]{dermer1992}
Dermer, C., Schlickeiser, R., \& Mastichiadis, A. 1992, Astronomy and Astrophysics (ISSN 0004-6361), vol. 256, no. 2, p. L27-L30., 256, L27

\bibitem[{Ding {et~al.}(2019)Ding, Gu, Geng, Xiong, Xue, Wang, \& Guo}]{ding2019}
Ding, N., Gu, Q., Geng, X., {et~al.} 2019, The Astrophysical Journal, 881, 125

\bibitem[{Donnarumma \& Vercellone(2019)}]{donnarumma2019}
Donnarumma, I., \& Vercellone, S. 2019, Rendiconti Lincei. Scienze Fisiche e Naturali, 30, 225

\bibitem[{Fan {et~al.}(2016)Fan, Yang, Liu, Luo, Lin, Yuan, Xiao, Zhou, Hua, \& Pei}]{fan2016}
Fan, J., Yang, J., Liu, Y., {et~al.} 2016, The Astrophysical Journal Supplement Series, 226, 20

\bibitem[{Fan {et~al.}(2018)Fan, Li, Liao, Chen, Liu, Lu, Yan, Zhang, Guo, Wu, {et~al.}}]{fan2018}
Fan, X.-L., Li, S.-K., Liao, N.-H., {et~al.} 2018, The Astrophysical Journal, 856, 80

\bibitem[{Fang {et~al.}(2022)Fang, Chen, Zhang, \& Wu}]{fang2022}
Fang, Y., Chen, Q., Zhang, Y., \& Wu, J. 2022, The Astrophysical Journal, 933, 224

\bibitem[{{Fermi Science Support Development Team}(2019)}]{2019ascl}
{Fermi Science Support Development Team}. 2019, {Fermitools: Fermi Science Tools}, Astrophysics Source Code Library, record ascl:1905.011

\bibitem[{Finke \& Becker(2014)}]{finke2014}
Finke, J.~D., \& Becker, P.~A. 2014, The Astrophysical Journal, 791, 21

\bibitem[{Foschini {et~al.}(2010)Foschini, Tagliaferri, Ghisellini, Ghirlanda, Tavecchio, \& Bonnoli}]{foschini2010}
Foschini, L., Tagliaferri, G., Ghisellini, G., {et~al.} 2010, Monthly Notices of the Royal Astronomical Society, 408, 448

\bibitem[{Gaidos {et~al.}(1996)Gaidos, Akerlof, Biller, Boyle, Breslin, Buckley, Carter-Lewis, Catanese, Cawley, Fegan, {et~al.}}]{gaidos1996}
Gaidos, J.~A., Akerlof, C.~W., Biller, S., {et~al.} 1996, Nature, 383, 319

\bibitem[{Ghisellini {et~al.}(2009{\natexlab{a}})Ghisellini, Nardini, Ghirlanda, \& Celotti}]{ghisellini2009}
Ghisellini, G., Nardini, M., Ghirlanda, G., \& Celotti, A. 2009{\natexlab{a}}, Monthly Notices of the Royal Astronomical Society, 393, 253

\bibitem[{Ghisellini \& Tavecchio(2008)}]{ghisellini2008}
Ghisellini, G., \& Tavecchio, F. 2008, Monthly Notices of the Royal Astronomical Society: Letters, 386, L28

\bibitem[{Ghisellini \& Tavecchio(2009)}]{ghisellini2009can}
---. 2009, Monthly Notices of the Royal Astronomical Society, 397, 985

\bibitem[{Ghisellini {et~al.}(2009{\natexlab{b}})Ghisellini, Tavecchio, Bodo, \& Celotti}]{ghisellini2009tev}
Ghisellini, G., Tavecchio, F., Bodo, G., \& Celotti, A. 2009{\natexlab{b}}, Monthly Notices of the Royal Astronomical Society: Letters, 393, L16

\bibitem[{Ghisellini {et~al.}(2011)Ghisellini, Tavecchio, Foschini, \& Ghirlanda}]{ghisellini2011}
Ghisellini, G., Tavecchio, F., Foschini, L., \& Ghirlanda, G. 2011, Monthly Notices of the Royal Astronomical Society, 414, 2674

\bibitem[{Hovatta \& Lindfors(2019)}]{hovatta2019}
Hovatta, T., \& Lindfors, E. 2019, New Astronomy Reviews, 87, 101541

\bibitem[{Kelly {et~al.}(2014)Kelly, Becker, Sobolewska, Siemiginowska, \& Uttley}]{kelly2014}
Kelly, B.~C., Becker, A.~C., Sobolewska, M., Siemiginowska, A., \& Uttley, P. 2014, The Astrophysical Journal, 788, 33

\bibitem[{Li {et~al.}(2018)Li, Hu, Wiita, \& Gupta}]{li2018}
Li, Y., Hu, S., Wiita, P.~J., \& Gupta, A.~C. 2018, Monthly Notices of the Royal Astronomical Society, 478, 172

\bibitem[{Li {et~al.}(2016)Li, Hu, Jiang, Chen, Priyadarshi, Li, Guo, \& Guo}]{li2016}
Li, Y.~T., Hu, S.~M., Jiang, Y., {et~al.} 2016, Publications of the Astronomical Society of the Pacific, 129, 014101

\bibitem[{Mannheim \& Biermann(1992)}]{mannheim1992}
Mannheim, K., \& Biermann, P. 1992, Astronomy and Astrophysics (ISSN 0004-6361), vol. 253, no. 2, Jan. 1992, p. L21-L24., 253, L21

\bibitem[{Marscher \& Travis(1996)}]{marscher1996}
Marscher, A., \& Travis, J. 1996, Astronomy and Astrophysics Supplement, v. 120, p. 537-540, 120, 537

\bibitem[{Meyer {et~al.}(2019)Meyer, Scargle, \& Blandford}]{meyer2019}
Meyer, M., Scargle, J.~D., \& Blandford, R.~D. 2019, The Astrophysical Journal, 877, 39

\bibitem[{Mukherjee {et~al.}(2019)Mukherjee, Mitra, \& Chatterjee}]{mukherjee2019}
Mukherjee, S., Mitra, K., \& Chatterjee, R. 2019, Monthly Notices of the Royal Astronomical Society, 486, 1672

\bibitem[{Nakagawa \& Mori(2013)}]{nakagawa2013}
Nakagawa, K., \& Mori, M. 2013, The Astrophysical Journal, 773, 177

\bibitem[{Nalewajko(2013)}]{nalewajko2013}
Nalewajko, K. 2013, Monthly Notices of the Royal Astronomical Society, 430, 1324

\bibitem[{Nandikotkur {et~al.}(2007)Nandikotkur, Jahoda, Hartman, Mukherjee, Sreekumar, B{\"o}ttcher, Sambruna, \& Swank}]{nandikotkur2007}
Nandikotkur, G., Jahoda, K.~M., Hartman, R., {et~al.} 2007, The Astrophysical Journal, 657, 706

\bibitem[{Newville {et~al.}(2016)Newville, Stensitzki, Allen, Rawlik, Ingargiola, \& Nelson}]{newville2016}
Newville, M., Stensitzki, T., Allen, D.~B., {et~al.} 2016, Astrophysics Source Code Library, ascl

\bibitem[{Nied{\'z}wiecki {et~al.}(2012)Nied{\'z}wiecki, Xie, \& Zdziarski}]{niedzwiecki2012}
Nied{\'z}wiecki, A., Xie, F.-G., \& Zdziarski, A.~A. 2012, Monthly Notices of the Royal Astronomical Society, 420, 1195

\bibitem[{Padovani {et~al.}(2019)Padovani, Oikonomou, Petropoulou, Giommi, \& Resconi}]{padovani2019}
Padovani, P., Oikonomou, F., Petropoulou, M., Giommi, P., \& Resconi, E. 2019, Monthly Notices of the Royal Astronomical Society: Letters, 484, L104

\bibitem[{Raiteri {et~al.}(2017)Raiteri, Villata, Acosta-Pulido, Agudo, Arkharov, Bachev, Baida, Ben{\'\i}tez, Borman, Boschin, {et~al.}}]{raiteri2017}
Raiteri, C.~M., Villata, M., Acosta-Pulido, J., {et~al.} 2017, Nature, 552, 374

\bibitem[{Roy {et~al.}(2019)Roy, Chatterjee, Joshi, \& Ghosh}]{roy2019}
Roy, N., Chatterjee, R., Joshi, M., \& Ghosh, A. 2019, Monthly Notices of the Royal Astronomical Society, 482, 743

\bibitem[{Ruan {et~al.}(2012)Ruan, Anderson, MacLeod, Becker, Burnett, Davenport, Ivezi{\'c}, Kochanek, Plotkin, Sesar, {et~al.}}]{ruan2012}
Ruan, J.~J., Anderson, S.~F., MacLeod, C.~L., {et~al.} 2012, The Astrophysical Journal, 760, 51

\bibitem[{Rulten(2022)}]{rulten2022}
Rulten, C. 2022, Galaxies, 10, 61

\bibitem[{Ryan {et~al.}(2019)Ryan, Siemiginowska, Sobolewska, \& Grindlay}]{ryan2019}
Ryan, J.~L., Siemiginowska, A., Sobolewska, M., \& Grindlay, J. 2019, The Astrophysical Journal, 885, 12

\bibitem[{Safna {et~al.}(2020)Safna, Stalin, Rakshit, \& Mathew}]{safna2020}
Safna, P., Stalin, C., Rakshit, S., \& Mathew, B. 2020, Monthly Notices of the Royal Astronomical Society, 498, 3578

\bibitem[{Saito {et~al.}(2013)Saito, Tanaka, Takahashi, Madejski, D’Ammando, {et~al.}}]{saito2013}
Saito, S., Tanaka, Y., Takahashi, T., {et~al.} 2013, The Astrophysical Journal Letters, 766, L11

\bibitem[{Sarkar {et~al.}(2019)Sarkar, Chitnis, Gupta, Gaur, Patel, Wiita, Volvach, Tornikoski, Chamani, Enestam, {et~al.}}]{sarkar2019}
Sarkar, A., Chitnis, V., Gupta, A., {et~al.} 2019, The Astrophysical Journal, 887, 185

\bibitem[{Scargle {et~al.}(2013)Scargle, Norris, Jackson, \& Chiang}]{scargle2013}
Scargle, J.~D., Norris, J.~P., Jackson, B., \& Chiang, J. 2013, The Astrophysical Journal, 764, 167

\bibitem[{Scarpa \& Falomo(1997)}]{scarpa1997}
Scarpa, R., \& Falomo, R. 1997, Astronomy and Astrophysics, v. 325, p. 109-123, 325, 109

\bibitem[{Schlickeiser(2009)}]{schlickeiser2009}
Schlickeiser, R. 2009, Monthly Notices of the Royal Astronomical Society, 398, 1483

\bibitem[{Sikora {et~al.}(1994)Sikora, Begelman, \& Rees}]{sikora1994}
Sikora, M., Begelman, M.~C., \& Rees, M.~J. 1994, The Astrophysical Journal, 421, 153

\bibitem[{Singal {et~al.}(2012)Singal, Petrosian, \& Ajello}]{singal2012}
Singal, J., Petrosian, V., \& Ajello, M. 2012, The Astrophysical Journal, 753, 45

\bibitem[{Sobolewska {et~al.}(2014)Sobolewska, Siemiginowska, Kelly, \& Nalewajko}]{sobolewska2014}
Sobolewska, M.~A., Siemiginowska, A., Kelly, B.~C., \& Nalewajko, K. 2014, The Astrophysical Journal, 786, 143

\bibitem[{Vaughan {et~al.}(2003)Vaughan, Edelson, Warwick, \& Uttley}]{vaughan2003}
Vaughan, S., Edelson, R., Warwick, R., \& Uttley, P. 2003, Monthly Notices of the Royal Astronomical Society, 345, 1271

\bibitem[{Wang {et~al.}(2022)Wang, Fan, Xiao, \& Cai}]{wang2022}
Wang, G., Fan, J., Xiao, H., \& Cai, J. 2022, Publications of the Astronomical Society of the Pacific, 134, 104101

\bibitem[{Xiong {et~al.}(2020)Xiong, Bai, Fan, Yan, Gu, Fan, Mao, Ding, Xue, \& Yi}]{xiong2020}
Xiong, D., Bai, J., Fan, J., {et~al.} 2020, The Astrophysical Journal Supplement Series, 247, 49

\bibitem[{Yang {et~al.}(2022)Yang, Fan, Liu, Tuo, Pei, Yang, Yuan, He, Wang, Wang, {et~al.}}]{yang2022}
Yang, J., Fan, J., Liu, Y., {et~al.} 2022, The Astrophysical Journal Supplement Series, 262, 18

\bibitem[{Yuan {et~al.}(2017)Yuan, Fan, Tao, Qian, Costantin, Xiao, Pei, \& Lin}]{yuan2017}
Yuan, Y.-H., Fan, J.-h., Tao, J., {et~al.} 2017, Astronomy \& Astrophysics, 605, A43

\bibitem[{Zhang {et~al.}(2023)Zhang, Tang, Wang, Wu, Jin, Dai, \& Zhu}]{zhang2023}
Zhang, B.-K., Tang, W.-F., Wang, C.-X., {et~al.} 2023, Monthly Notices of the Royal Astronomical Society, 519, 5263

\bibitem[{Zhang {et~al.}(2022{\natexlab{a}})Zhang, Zhao, \& Wu}]{zhangchu2022}
Zhang, B.-K., Zhao, X.-Y., \& Wu, Q. 2022{\natexlab{a}}, The Astrophysical Journal Supplement Series, 259, 49

\bibitem[{Zhang {et~al.}(2022{\natexlab{b}})Zhang, Yan, \& Zhang}]{zhang2022}
Zhang, H., Yan, D., \& Zhang, L. 2022{\natexlab{b}}, The Astrophysical Journal, 930, 157

\end{thebibliography}
\bibliographystyle{aasjournal}

\end{document}